\documentstyle[12pt]{article}
\begin{document}

\renewcommand{\thesection}{\arabic{section}.}
\renewcommand{\thesubsection} {\arabic{subsection}.}
\renewcommand{\theequation}{\arabic{equation}}
\renewcommand {\c}  {\'{c}}
\newcommand {\cc} {\v{c}}
\newcommand {\s}  {\v{s}}
\newcommand {\CC} {\v{C}}
\newcommand {\C}  {\'{C}}
\newcommand {\Z}  {\v{Z}}

%
\begin{center}
{\bf On infinite quon statistics and "ambiguous" statistics
 }
\\[8.0mm]

S.Meljanac$^{+}$  \\
 Rudjer Boskovic  Institute, Bijenicka c.54, 10001 Zagreb,
Croatia\\[3mm] 
$^{+}$ E-mail: meljanac@thphys.irb.hr\\[7mm]

M.Milekovic $^{++}$\\
 Prirodoslovno-Matematicki Fakultet, Zavod za teorijsku fiziku, 
 Bijenicka c.32,
\\ 10000 Zagreb, Croatia \\[3mm] 
$^{++}$ E-mail: marijan@phy.hr\\[7mm]

R.Ristic$^{+++}$\\
 Faculty of Education, L.Jagera 9,\\
  31 000 Osijek, Croatia\\[3mm]
$^{+++}$ E-mail: rristic@knjiga.pedos.hr

\end{center}
\setcounter{page}{1}
\bigskip


\vskip 0.2cm
We critically examine a recent suggestion that 
 "ambiguous" statistics is equivalent to infinite quon statistics and that it describes 
 a dilute, nonrelativistics ideal gas of extremal black holes. 
We show that  these two types of statistics are different and that the description of
extremal black holes in terms of "ambiguous" statistics cannot be applied.

\newpage

\pagestyle{plain}
\def\leer{\vspace{5mm}}
\baselineskip=24pt
\setcounter{equation}{0}

${\bf {1.  Introduction.}}\\
$A few years ago, it was argued $^{1}$ that  extremal black holes obeyed infinite quon 
statistics , ${\it i.e.}$, that  quantum states of extremal black holes belonged to the (quantum)
 Boltzmannian space $^{2,3}$. 
 This connection of extremal black holes to infinite quon statistics was later adopted by several authors $^{4,5}$ 
 and, recently, the thermodynamics of the ensemble of extremal black holes was discussed from the point of view of 
 statistical mechanics of  "ambiguous" statistics $^{6}$, which interpolates between Bose and Fermi 
 statistics and was  claimed to be equvalent to quon statistics.\\
 The aim of this paper is to closely examine  the supposed connection between 
  "ambiguous" statistics and infinite quon statistics and their relation to the statistics of extremal black holes.
   First, we briefly review  quon statistics and the corresponding thermodynamics. We mention a few 
  shortcomings of the  thermodynamics of  quonic systems. Then, we show that "ambiguous" statistics  
is different from  quon statistics and that there is no operator realization of it. 


${\bf {2. Quon \, algebra.}}$\\
 Recall that a free system of particles obeying infinite quon statistics 
is described by the following commutation relations $^2$ (existence of the unique vacuum 
state $|0>$ is assumed ):\\
\begin{equation}
a_{i}a_{j}^{\dagger} - qa_{j}^{\dagger}a_{i} = \delta_{ij},
\end{equation}
$$
a_{i}|0> =0 \qquad <0|0> =1.
$$
Here $i,j$ are discrete indices,$i,j = 1,2,...M$, and the parameter $q$ is a 
real number, $q \in {\bf R}$.\\
 The main properties of quons are as follows:\\
 (i)  Norms of the states are positive definite for $-1< q < 1$.\\
 (ii) For $q^{2}\neq 1$, the commutation relations do not exist
between annihilation (creation) operators $a_{i}$,$a_{j}$
($a_{i}^{\dagger}$,$a_{j}^{\dagger}$), ${\it i.e.}$, there are $n!$ linearly
independent states $a_{i_{1}}^{\dagger}\dots a_{i_{n}}^{\dagger}|0>$ for different
permutations of fixed indices $1,2,\dots ,n$.\\
 (iii) The number operator exists in the form of an infinite
series expanded in powers of creation and annihilation operators.\\
 (iv) The particles obeying infinite quon statistics do not have a local-field theory, there is no
 spin-statistics restriction for such particles and they can have any spin. Nevertheless, the 
  TCP theorem and the clustering property is valid for free infinite-statistics fields.

${\bf {3. Statistical \,  mechanics \, of \,  the \, quon \, gas.}}$\\
 To discuss the statistical mechanics of the 
quon gas, we assume that  quon particles are massive (with mass $m$), 
 spinless, nonrelativistic and noninteracting. The Hamiltonian of an ensemble of  particles 
is that of the free system :
\begin{equation}
H=\sum_{k=1}^{M} E_k \, N_k \, ,
\end{equation}
\noindent
where $E_k$ is the energy of the $k^{th}$ level and $N_k$ are the number operators 
counting particles on the $k^{th}$ level. 
Recall that the statistical average of an 
observable,described by an operator ${\cal {O}}$ in a given ensemble, is defined as
\begin{equation}
 <{\cal O}>=\frac{Tr  {\cal O}\,e^{-\beta H}}{Tr e^{-\beta H}}=\frac{1}{{\cal Z}}
 Tr {\cal O} e^{-\beta H}.
\end{equation}
Here ${\cal Z}$ is the thermodynamical (grand-canonical) partition function for a multi-level system described by 
M independent creation (annihilation) operators $a^{\dagger}_{i}$ ( $a_i$ ), \\$i=1,2,..M.$
 The partition function ${\cal Z}$ for a free system 
described by the algebra, Eq.(1), is a power series in $ x_k=e^{\beta \mu - \beta E_k}$  
( where $\mu $ denotes a chemical potential ) and is given by (compare Ref.(7))
\begin{eqnarray}
{\cal Z}(x_1,...x_M) & = &1 + \sum_{k=1}^M x_k + ( \sum_{k=1}^M x_k )^2 + \cdots 
= 1+ z_1 + ( z_1 )^2 + \cdots \\ \nonumber
 & = & \frac{1}{1-z_1}, 
\end{eqnarray}
for $z_1 = \sum_{k=1}^M x_k < 1$ . Notice that the partition function does not 
depend on the interpolation parameter $q$, which is a generic property of all generalized 
statistics in the second quantized approach. From ${\cal Z}(x_1,...x_M) $ one can derive, e.g., distribution 
function $n_k(E_k)$ ( occupation number of the k-th level)   and an average particle number $< N >$ as
\begin{equation}
n_k(E_k)=x_k \frac{\partial}{\partial x_k} ln \,{\cal Z}(x_1,...x_M) = \frac{x_k}{1 - z_1},
\end{equation}
$$
< N > = \frac{1}{\beta}\frac{d}{d \mu} ln \,{\cal Z}(x_1,...x_M)= \frac{z_1}{1 - z_1}.
$$
In the limit $M \rightarrow \infty $ and $E_k = E = \frac{p^2}{2m}$, transforming 
$ 
\sum_k \rightarrow \int \frac{d\, V}{(2\pi \hbar )^3} d^3p ,
$
 one recovers the partition functions 
used in Refs.(8,9), namely,
\begin{eqnarray}
(z_1)_{\infty} &=& e^{\beta \mu }\,(\frac{V}{\lambda ^3})\\ \nonumber
({\cal Z})_{\infty} &=& \sum_{N=0}^{\infty}( (z_1)_{\infty} )^N
\end{eqnarray}
where
 $\lambda =  (\frac{2\pi \beta \hbar ^2}{m})^{\frac{1}{2}}$ is the thermal de Broglie wavelength. 
 In this limit, Eq.(5) yields
 $$
< N >_{\infty}= \frac{e^{\beta \mu } V}{\lambda^3 -e^{\beta \mu } V}.
 $$
 
 It is well known $^{8,9}$ that quon gas, described by Eq.(6), is plagued with several 
 difficulties. The most serious ones are:\\
 (1) The partition functions (6) exhibit Gibbs' paradox which cannot be resolved by 
 fixing the overall statistical weight of the $N$-particle phase space as in  classical statistical 
 mechanics, since the weight is unambiguously determined by quon 
 quantum-statistical mechanics. \\
 (2) the entropy $S$ or an average particle number  fails to be an extensive variable of the system; 
 moreover, they both diverge 
 for the critical volume $V= e^{-\beta \mu }\lambda ^3$.

The pathological behavior of the free quon gas can also be  detected for  finite $M$. To show this, 
we apply  the usual method of Lagrange's undetermined multipliers $^10$ in order  to extremize the entropy $S$ 
under subsidiary conditions ( constancy of the total number of particles, N , and the total 
energy of the gas, E ):
\begin{equation}
\frac{\partial}{\partial N_k } ( S - \alpha \sum_{k=1}^{M} N_k -\beta  \sum_{k=1}^{M}E_k N_k ) = 0,
\end{equation}
$$
\sum_{k=1}^{M} N_k = N , \qquad \sum_{k=1}^{M}E_k N_k =E.
$$
Here, the entropy $S$ is related to the number of all allowed N-body states distributed 
over M energy levels, $W(M,N)$, as $ S = ln \, W(M,N) $. A simple counting procedure, in 
agreement with (1), gives for $W(M,N)$ 
 $$
 W(M,N)= \sum_{\sum_{k=1}^M N_k = N} \frac{N!}{N_1! N_2! \cdots N_M!} = M^N.
 $$ 
 Solving Eq.(7) gives a constraint
\begin{equation}
\beta E_k = ln M - \alpha = const , \qquad \forall k=1,2,..M.
\end{equation}
In such a thermal equilibrium state , after identifying $\alpha = - \beta \mu$, one would have
$ x_k = \frac{1}{M}$ and  $ z_1 = 1 $. Hence, the partition function diverges, $ {\cal Z}(x_1,...x_M)
\rightarrow \infty $, and, as a consequence, quantities such as  $< N >_{\infty} $   also diverge.
The conclusion is obvious. Regardless of finite or infinite $M$ and for any 
$-1< q < 1$, the thermodynamical limit does not make sense for the free quon gas. 

${\bf {3. "Ambiguous" \,\, statistics.}}$ \\
The basic idea of  "ambiguous" statistics $^6$ is that only primary Bose and 
Fermi statistics are allowed, but the statistics of the  particle may fluctuate between Bose and Fermi statistics 
and is determined 
in a probabilistic way. During the interaction, the first particle identifies the second one as either a boson or a fermion,
 with  respective  
 probabilities $p_b$ and  $p_f$. The system of $N$ 
particles effectively consists of $k$ bosons and $N - k$ fermions and the probability of this 
N-particle realization is $p_b^k\,p_f^{N-k}$. The total number of  states of the system 
is
\begin{equation}
W (G_j,N_{j})=\Pi_{j} \, \sum_{k=0}^{N_j}\left(\stackrel{\displaystyle N_j}{k} \right)
\left(\stackrel{\displaystyle G_j + k -1}{k} \right)
\left(\stackrel{\displaystyle G_j }{N_j -k} \right)
p_b^kp_f^{N_j-k},
\end{equation}
where $G_j$ is the number of allowed one-particle states of type $j$.  
It is claimed in Ref.(6) that the operator realization of this statistics is described by Eq.(1), 
with the deformation parameter 
$q=\frac{p_b - p_f}{p_b + p_f}$. Particularly, for $p_b = p_f$,
 one should recover Greenberg's infinite statistics with $q=0$ (quantum Boltzmann statistics). 
 We have several objections to this 
 identification.\\
Indeed, from  Eq.(9), one
 can derive the distribution 
function $n_j$ and the thermodynamic properties by varying the  entropy, as in Eq.(7). However, from the 
 definition (9) alone, one cannot determine the operator algebra ( commutation relation ) for the 
 annihilation and creation operators. First of all, it is impossible to obtain the fractional 
 number of states, Eq.(9),  in  Fock space. Whether or not there exists an operator algebra, leading 
 to the same thermodynamic properties as for "ambiguous" statistics, is an open question. Furthermore, 
 we point out that the proposed 
 "ambiguous" statistics is completely different from  infinite statistics. For the case  $p_b = p_f\equiv p$, 
 the distribution function $n_j$, derived from Eqs.(7,9), can be rewritten as 
 $$
N_j=\frac{2p}{ e^{\beta (  E_j - \mu )} + p} + \frac{2p}{ e^{\beta (  E_j - \mu )} - p} \equiv N_j(p) +N_j(-p)
 $$
and has nothing to do with infinite statistics. However, $N_j(\pm p)$ is exactly the distribution derived 
from the counting rule
$$
W_j(N_j)=\frac{1}{N_j!}( \frac{(G_j)!}{(G_j - p\, N_j)!} )^{\frac{1}{p}}. 
$$ 
The thermodynamics of  such a system is described in Ref.(11).\\ 
 In defining "ambiguous" statistics, 
 two unjustified assumptions are made:\\
 (i) First, that the annihilation and creation operators satisfy Bose ($q=-1)$, Fermi ($q=+1)$, ${\it and}$  quon ($q=0$) 
 statistics:
 $$
 a_{i}a_{j}^{\dagger} - a_{j}^{\dagger}a_{i} = \delta_{ij}, 
$$ 
$$
a_{i}a_{j}^{\dagger} + a_{j}^{\dagger}a_{i} = \delta_{ij},
$$
$$
a_{i}a_{j}^{\dagger}= \delta_{ij},
$$
with the probabilities $p_b^2$, $p_f^2$, and $2p_bp_f$, respectively. 
This is in contradiction with the initial assumption that ${\it only}$ Bose and Fermi statistics are allowed.\\
(ii) Second, in the second quantized approach, it is not mathematically correct $^{12}$ to add the above commutation 
relation with relative weights
$p_b^2$, $p_f^2$, and $2p_bp_f$ in order 
to obtain Eq.(1)
with  the deformation parameter $q=\frac{p_b - p_f}{p_b + p_f}$. Instead, one can postulate Eq.(1) ${\it ab}$
 ${\it initio}$ and 
investigate the thermodynamics of particles obeying quon statistics but this thermodynamics is completely different from 
that studied in Ref.(6). Hence, we may conclude that "ambiguous" statistics, as defined in Ref.(6), possesses no Fock-like 
realization and  is completely different from infinite quon statistics.

We mention that a similar idea,  ${\it i.e.}$, that, basically, only symmetric (${\it i.e.}$, Bose-like) and 
antisymmetric (${\it i.e.}$, Fermi-like) IRREP's of the permutation group $S_N$ are allowed 
in generalized statistics, was suggested and investigated in Ref.(13). 
In this approach,  generalized statistics is described by an algebra 
for which there exists a Fock - like realization. The construction is based on two vacuua $|\pm >$ and on 
the commutation rules containing a unitary operator $\hat{q}$:  
\begin{equation}
a_{i}a_{j}^{\dagger} - \hat{q} a_{j}^{\dagger}a_{i} = \delta_{ij}, \qquad \hat{q}|\pm > = \pm 1 |\pm >.
\end{equation}
 The corresponding Fock - like representations are built on a linear combination of  $|\pm >$  vacuua. 
 In Ref.(14) the minimal generalized 
statistics with permutation-group invariance, interpolating between Bose and Fermi statistics, is constructed.
Moreover, it is shown that $a_{i}a_{j}^{\dagger}$ can be brought to a normally ordered form without using the 
 operator $\hat{q}$ and that a complete theory can be formulated using a single vacuum state $|0>$ (instead of 
$|\pm >$ vacuua). We notice that this type of statistics is also different from  infinite quon statistics.

${\bf {4. Conclusion.}}$ \\
"Ambiguous" statistics, defined through the counting rule (9), although interesting as a 
possible generalization of Bose and Fermi statistics, is completely different from infinite statistics, and 
Eq.(1) does not represent its operator realization. If one seriously takes  the suggestion that extremal black holes 
exhibit the $q=0$ quonic behavior, then "ambiguous" statistics, not being equivalent to quon statistics,  does not 
describe extremal black holes. Moreover,   the assumption  that $q=0$ quons may be viewed as distinguishable particles with infinite internal 
degrees of freedom would also lead to difficulties and would not resolve Gibbs' paradox $^8$. We believe that a correct statistics 
of extremal black holes is still an interesting open question.

 \newpage
\baselineskip=24pt
{\bf References}
\begin{description}

\item{1.}
A.Strominger, Phys.Rev.Lett. {\bf 71}, 3397 (1993)  .
\item{2.}
 O.W.Greenberg , Phys.Rev.Lett. {\bf 64}, 705 (1990)  ;
 S.Meljanac  and A.Perica , Mod.Phys.Lett. {\bf A9}, 3293 (1994)  ;
J.Phys.{\bf A27}, 4737 (1994).  
\item{3.}
A.B.Govorkov , Phys.Part.Nucl.{\bf 24},565 (1993) .
\item{4.}
D.Minic , hep-th/9712202.
\item{5.}
I.V. Volovich , hep-th/9608137. 
\item{6.}
M.V. Medvedev ,  Phys.Rev.Lett.{\bf 78}, 4147 (1997)  .
\item{7.} 
S.B. Isakov , Int.J.Theor.Phys. {\bf 32}, 737  (1993)  .
\item{8.} 
R.F.Werner , Pys.Rev.{\bf D48},  2929  (1993)  .
\item{9.}
J.W.Goodison  and D.J.Toms , Phys.Rev.Lett. {\bf 71},  3240 (1993)  .
\item{10.}
L.D.Landau  and E.M.Lifshitz , {\it Statistical Physics} ( Pergamon Press, Oxford ) 1969.
\item{11.}
K. Byczuk et al., Acta Phys.Pol. {\bf 26},  2167 (1995)  .
\item{12.} 
R.Speicher , Lett.Math.Phys. {\bf 27},  97  (1993)  ; O.W. Greenberg , hep-ph/9306225.
\item{13.}
L.Wu and  Z.Wu , Phys.Lett. {\bf A170},  280  (1992)  ; R. Scipioni , Phys.Lett. {\bf B327},  56  (1994); 
Nuovo Cimento {\bf B109},  479 (1994) .
\item{14.}
B. Melic  and  S.Meljanac , Phys.Lett. {\bf A226},  22  (1996) .

\end{description}
\end{document}